\documentclass[aps,prl,twocolumn,superscriptaddress,nofootinbib]{revtex4-2}

\usepackage{amsmath,amssymb}
\usepackage{graphicx}
\usepackage{braket}
\usepackage{hyperref}
\usepackage{physics}

\begin{document}

\title{Probing graph topology from local quantum  measurements}

\author{Francesco Romeo}
\affiliation{Dipartimento di Fisica ”E. R. Caianiello”,
Università degli Studi di Salerno,
Via Giovanni Paolo II, I-84084 Fisciano (SA), Italy}
\affiliation{INFN, Gruppo Collegato di Salerno, Via Giovanni Paolo II, I-84084 Fisciano (SA), Italy}

\author{Jacopo Settino}
\affiliation{Dipartimento di Fisica, Università della Calabria, Rende (CS), I-87036, Italy}
\affiliation{INFN, Gruppo Collegato di Cosenza, I-87036 Arcavacata di Rende (CS), Italy}

\begin{abstract}
We show that global properties of an unknown quantum network, such as the average degree, hub density, and the number of closed paths of fixed length, can be inferred from strictly local quantum measurements. In particular, we demonstrate that a malicious agent with access to only a small subset of nodes can initialize quantum states locally and, through repeated short-time measurements, extract sensitive structural information about the entire network. The intrusion strategy is inspired by extreme learning and quantum reservoir computing and combines short-time quantum evolution with a non-iterative linear readout with trainable weights. These results suggest new strategies for intrusion detection and structural diagnostics in future quantum Internet infrastructures.
\end{abstract}
\maketitle

\paragraph*{Introduction.---} 
The quantum Internet promises to enable secure communication, distributed quantum computation, and large-scale entanglement sharing by interconnecting quantum devices across complex physical networks \cite{QI1,QI2,QI3,QI4,QI5}. Its design typically relies on the assumption that the network topology is either known or controlled and that security stems primarily from quantum cryptographic primitives such as key distribution and authentication protocols \cite{cry0,cry1,cry1.1,cry2,cry3,cry4}. However, a largely unexplored vulnerability arises from the inherent nature of quantum dynamics: a malicious agent with local access to a small region of the network may exploit coherent evolution and measurement to extract sensitive information about the global structure.

This form of \textit{quantum reconnaissance} is qualitatively different from classical probing and opens a new front in the security analysis of quantum networks. Unlike classical diffusion processes\cite{DiffCl}, short-time quantum evolution \cite{QW2,QW3,QW5} exploits coherent superpositions and interference effects, resulting in a richer nonlinear dependence on the adjacency matrix. This enhanced expressivity enables more effective structural inference, even under severely constrained access. Existing proposals in quantum machine learning, such as quantum reservoir computing \cite{QREX1,QREX2,QREX4, QREX5,QREX6,QREX9,QREX11,QREX12,QREX15},  extreme learning architectures \cite{QREX3,QREX8,QREX10,QREX13}, and quantum reservoir probing \cite{QREX7,QREX14}, have demonstrated the expressivity of quantum systems as learning substrates, yet their application to network structure inference\cite{NetInference1,NetInference2,NetInference3} in adversarial scenarios remains unaddressed.

\begin{figure}[t!]
    \centering
    \includegraphics[width=0.48\textwidth]{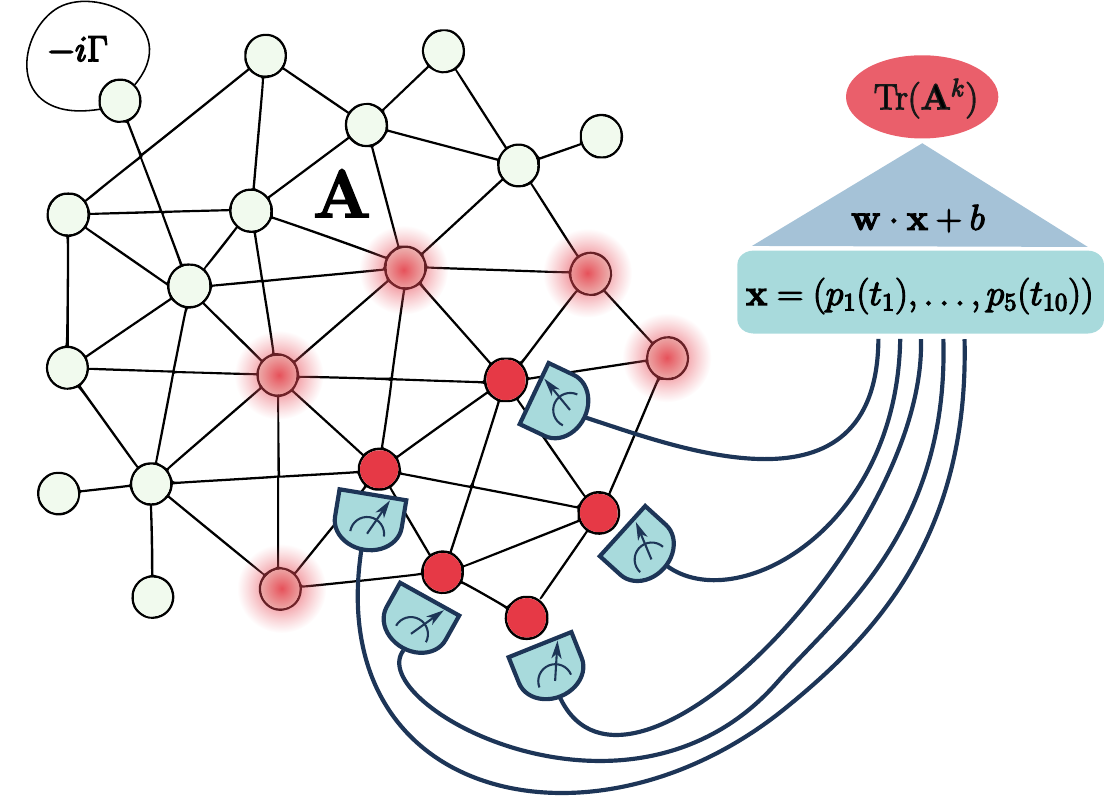}
    \caption{Schematic of the learning protocol. A quantum excitation is initialized on a subset of accessible nodes (red) within a larger network of unknown structure encoded by the adjacency matrix $A$ and let evolve. Site occupations $p_i(t_k)$ on the monitored nodes are recorded at successive time steps. The resulting input vector $\boldsymbol{x}$ feeds a linear readout trained to infer global observables such as $\mathrm{Tr}(A^k)$, hub density, network size, and information leakage parameter $\Gamma$. The method enables inference of the network topology from local quantum measurements.}
    \label{fig:fig1}
    \vspace{-0.55cm}
\end{figure}

In this work, we introduce a protocol that demonstrates how key global properties, such as average degree, hub density, and cycle statistics, can be inferred by monitoring the short-time evolution of local observables on a small subset of nodes (Fig.~\ref{fig:fig1}). Our approach leverages the short-time quantum dynamics to extract nonlinear features from the network Hamiltonian, and uses a non-iterative linear readout with trainable weights to predict global graph observables. The same strategy can be employed by a legitimate monitoring agent to detect unauthorized modifications of the network structure. We show that parasitic connections, such as those resulting from the attachment of external subgraphs by a malicious intruder, can be modeled as localized non-Hermitian perturbations, whose presence is revealed by systematic deviations in the quantum dynamics. These results establish a dual perspective: they highlight both a novel quantum-native vulnerability in future networked architectures, and a viable diagnostic mechanism for intrusion detection based on minimal, local quantum measurements.

\paragraph*{Statement of the problem, model and assumptions.---} 
The central question addressed in this work is whether an agent with access to a limited subset of nodes in a quantum network can extract global information about the entire structure of the network. We suppose that the agent is able to prepare a local quantum excitation on the accessible sites and perform repeated measurements of local observables, with no prior knowledge of the network topology beyond those nodes. We ask to what extent this limited access can be leveraged to infer nonlocal structural properties of the network by monitoring the short-time quantum evolution of the produced excitation.

To avoid unnecessary technical complications, we adopt the single-particle Schr\"odinger framework, which captures the essential physics while remaining applicable to a broad class of quantum platforms, including photonic\cite{QWPhotons1,QWPhotons2}, spin\cite{QWSpin}, and superconducting qubit networks\cite{QWSqubit}.

Thus, to our purposes, we consider a quantum excitation evolving on an undirected network of $N$ nodes, represented by a real symmetric \emph{adjacency matrix} $A$. The entries of $A$ are defined as $A_{ij} = 1$ if nodes $i$ and $j$ are connected by an edge, and $A_{ij} = 0$ otherwise. As such, the degree of node $i$, defined as the number of its neighbors, is given by $d_i = \sum_j A_{ij}$, and global topological information (such as degree distributions, presence of hubs, or short cycles) is encoded in the spectrum and powers of $A$.
In fact, in spectral graph theory\cite{Chung1997}, global structural features are encoded in the spectral moments $\mathcal{M}_k=\operatorname{Tr}(A^k)$, with the integer $k > 1$. 
This follows from the fact that the matrix element $[A^k]_{ij}$ counts the number of paths of length $k$ connecting nodes $i$ and $j$, while $[A^k]_{ii}$ enumerates closed walks of length $k$ starting and ending at node $i$. 

The excitation undergoes unitary dynamics generated by the Hamiltonian

\begin{equation}
H = \gamma A,
\end{equation}
where $\gamma$ is an appropriate energy scale. The time evolution operator, associated with the Schr\"odinger
 equation $H|\psi\rangle =i\hbar \partial_t |\psi\rangle$, is
\begin{equation}
U(t) = e^{-i \gamma t A / \hbar},
\end{equation}
which controls a single-particle continuous-time quantum walk on the graph \cite{QW1,QW4,QW6,QW7,QW8,QW9,QW10,QW11}, according to the relation $|\psi(t)\rangle =U(t)|\psi(0)\rangle$, with $|\psi(0) \rangle$ the initial quantum state of the local excitation.

Let $\mathcal{S} = \{i_1, \dots, i_M\}$ denote the set of accessible nodes, with cardinality $|\mathcal{S}|\equiv M \ll N$, and let the system be prepared in an initial state $\ket{\psi(0)}$ uniformly distributed over $\mathcal{S}$. The state of the system is monitored at a sequence of short times $\{t_k\}_{k=1}^T$. At each time $t_k$, the local occupation probabilities 
\begin{equation}
p_i(t_k) = \langle \psi(t_k) | \Pi_i | \psi(t_k) \rangle,
\end{equation}
with $\Pi_i = |i\rangle \langle i|$ the projector on node $i \in \mathcal{S}$,
are estimated through repeated projective measurements at accessible sites. \\
We focus on \emph{short-time regime}, defined by the condition:
\[
\tau = \frac{\gamma t_T}{\hbar} < 1.
\]
In this regime, higher-order terms of the unitary evolution expansion, i.e.
\begin{equation}
U(t_T) = \mathbb{I} - i \tau A - \frac{1}{2} \tau^2 A^2 + \cdots,
\end{equation}
are increasingly suppressed by powers of $\tau$, so that $U(t_T)$ is predominantly influenced by the low-order powers of the adjacency matrix. The same argument holds for $t_k < t_T$. Furthermore, the expansion of $U(t_k)$, compared to that of $U(t_T)$, is dominated by lower powers of $A$, so that the dynamics probes distinct spectral features at each time step.
These observations imply that the short-time quantum evolution contains all the relevant information to infer nonlocal structural properties of a network.

A crucial assumption of the protocol is that the underlying network remains static throughout the measurement process, i.e., the matrix $A$ does not change while statistics is acquired. This condition is satisfied on physical platforms where the timescale of topological reconfiguration of the network is much longer than the total duration of sampling. Even in scenarios where the graph evolves slowly, we expect the inference strategy to remain rather effective, albeit with degraded accuracy due to fluctuations in the underlying structure.\\

\begin{table*}[t!]
\centering
\resizebox{\textwidth}{!}{%
\begin{tabular}{lccccc}
\hline\hline
Observable & Training Set (TrS) & Test Set (TS) & MAPE (TrS) & MAPE (TS) & $r \ \text{(TrS)} \ | \ r \ \text{(TS)}$ \\
\hline\hline
$Tr[A^2]$ & $[150\, G(50,0.6)]_{150}$ & $[50\, G(50,0.6)]_{50}$ & 0.88\% & 1.25\% & * \\
\hline
$Tr[A^2]$ & $[350\, G(50,0.6)]_{350}$ & $[50\, G(50,0.6)]_{50}$ & 1.05\% & 1.16\% & * \\
\hline
$Tr[A^2]$ & $[90\, G(50,p \in\{0.2,0.4,0.6,0.8\})]_{360}$ & $[10\, G(50,p \in\{0.2,0.4,0.6,0.8\})]_{40}$ & 5.63\% & 6.04\% & * \\
\hline\hline
$Tr[A^3]$ & $[150\, G(50,0.6)]_{150}$ & $[50\, G(50,0.6)]_{50}$ & 2.83\% & 3.75\% & * \\
\hline
$Tr[A^3]$ & $[350\, G(50,0.6)]_{350}$ & $[50\, G(50,0.6)]_{50}$ & 3.19\% & 3.86\% & * \\
\hline
$Tr[A^3]$ & $[90\, G(50,p \in\{0.2,0.4,0.6,0.8\})]_{360}$ & $[10\, G(50,p \in\{0.2,0.4,0.6,0.8\})]_{40}$ & 33.53\% & 34.97\% & $0.99484 \ | \ 0.99485$ \\
\hline\hline
$Tr[A^4]$ & $[150\, G(50,0.6)]_{150}$ & $[50\, G(50,0.6)]_{50}$ & 3.46\% & 4.79\% & * \\
\hline
$Tr[A^4]$ & $[350\, G(50,0.6)]_{350}$ & $[50\, G(50,0.6)]_{50}$ & 3.98\% & 4.66\% & * \\
\hline
$Tr[A^4]$ & $[90\, G(50,p \in\{0.2,0.4,0.6,0.8\})]_{360}$ & $[10\, G(50,p \in\{0.2,0.4,0.6,0.8\})]_{40}$ & 54.44\% & 59.19\% & $0.99535 \ | \ 0.99569$ \\
\hline\hline
Hub density & $[360\, G(100,0.5)]_{360}$ & $[40\, G(100,0.5)]_{40}$ & 9.55\% & 8.88\% & * \\
\hline
Hub density & $[90\, G(100,p \in\{0.2,0.4,0.6,0.8\})]_{360}$ & $[10\, G(100,p \in\{0.2,0.4,0.6,0.8\})]_{40}$ & 12.01\% & 10.34\% & * \\
\hline\hline
Network Size ($n$) & $[90\, G(n \in \{20,40,60,80\},0.5)]_{360}$ & $[10\, G(n \in \{20,40,60,80\},0.5)]_{40}$ & 7.05\% & 7.88\% & * \\
\hline
$|\lambda_{min}/\lambda_{max}|$ & $[90\, G(50,p \in\{0.2,0.4,0.6,0.8\})]_{360}$ & $[10\, G(50,p \in\{0.2,0.4,0.6,0.8\})]_{40}$ & 7.9\% & 8.22\% & * \\
\hline
Non-Hermitian Parameter ($\Gamma$) & $[G_\Gamma(100,0.5)]_{360}$ & [$G_\Gamma(100,0.5)]_{40}$ & 13.53\% & 2.14\% & $0.99867 \ | \ 0.99913$ \\
\hline\hline
\end{tabular}
}
\caption{Performance summary for representative learning tasks. The training (TrS) and test (TS) sets are composed of Erdős–Rényi graphs $G(n,p)$ without isolated nodes or disconnected subgraphs. A dataset (see ref. \cite{gitHubLink} for the code) composed of $m_1$ graphs of type $G(n_1,p_1)$, $m_2$ graphs of type $G(n_2,p_2)$, ..., $m_r$ graphs of type $G(n_r,p_r)$, is denoted by
\(
\left[m_1 G(n_1,p_1),\, m_2 G(n_2,p_2),\,\ldots,\, m_r G(n_r,p_r)\right]_{\sum_k m_k},
\)
where the subscript $\sum_k m_k$ indicates the total size of the dataset. The shortened notation 
\(
\left[m G(n,p \in \{p_1,...,p_r\})\right]_{m \cdot r},
\)
can be used, for example, when $m_1=m_2=...=m_r \equiv m$ and $n_1=n_2=...=n_r \equiv n$. A similar notation is introduced when the network size varies in a set, while the connection probability is fixed.
The model is trained to infer the observables from short-time quantum dynamics using ridge regression. The initial quantum excitation is equally distributed over five selected nodes, and local occupations $p_i(t_k)$ are sampled at discrete times $t_k = k\, \Delta t$ with $k \in \{1,\dots,10\}$. MAPE for the training and the test set, complemented when necessary by the Pearson correlation coefficient $r$, is reported. MAPE values are generally lower when training is performed on graphs belonging to same ensemble and tend to increase when the dataset includes structurally heterogeneous graphs.
}
\label{tab:results}
\end{table*}
 
\paragraph*{From local quantum evolution to graph inference.---} 
The occupation probabilities recorded from short-time quantum evolution encode a set of nonlinear functionals of the adjacency matrix $A$, implicitly reflecting global structural features of the network. We now describe how such local information can be processed to infer global observables through a simple, non-iterative learning strategy.

The $p_{i_k} (t_j)$ values are collected in a feature vector of dimension $MT$:
\[
\mathbf{x} = \left( p_{i_1}(t_1),\dots, p_{i_M}(t_1), \dots, p_{i_1}(t_T),  \dots, p_{i_M}(t_T) \right).
\]
To extract a given global observable $\mathcal{O}$ (e.g., average degree, hub density, cycle count), we define a linear estimator
\begin{equation}
\hat{\mathcal{O}} = \sum_{\mu=1}^{MT} w_\mu x_\mu + b,
\end{equation}
where the weights $\{w_\mu\}$ and the bias $b$ have been previously optimized for the specific task as detailed subsequently.

This protocol is inspired by the philosophy of \emph{Extreme Learning Machines (ELM)} and \emph{Quantum Reservoir Computing (QRC)}, where a fixed nonlinear transformation of the input is processed via a linear readout layer. In our case, short-time quantum evolution provides a non-trivial source of nonlinearity, eliminating the need for iterative optimization or random parameter initialization. The inferred global observable $\hat{\mathcal{O}}$ can be understood as an unknown functional of $A$ expressed as a linear combination of non-linear functionals, according to the relation:
\begin{equation}
\hat{\mathcal{O}}[A] = \sum_{\mu=1}^{MT} w_\mu\, \mathcal{F}_\mu[A] + b,
\end{equation}
where $\mathcal{F}_\mu[A]\equiv x_\mu$ denotes the $\mu$-th implicit nonlinear functional of the adjacency matrix, determined by the quantum evolution and local measurement process.

As stated above, the readout weights must be trained before their use on a representative set of graphs with known topology. These can be drawn from simulated graph models (e.g., Erdős–Rényi\cite{graph1,graph2,graph3}, Barabasi-Albert\cite{graph4}, Watts-Strogatz\cite{graph5}) or obtained from experiments on quantum networks of known topology. Once trained, the readout can be applied to infer structural properties of unseen networks using only local quantum measurements.
\\
\paragraph*{Training the model.---} 
We evaluated the performance of our protocol on various inference tasks by simulating short-time quantum dynamics on connected Erdős–Rényi graphs. Although this class is characterized by the absence of strong structural correlations, we have verified that similar results hold for alternative architectures such as scale-free and small-world networks.

We introduce the notation $G(n,p)$ to indicate a graph sampled from the Erdős–Rényi ensemble with $n$ nodes and independent edge probability $p$. Training and test datasets are obtained by ordered sequences of different graphs of the mentioned class. Graphs with disconnected components or isolated nodes are discarded \textit{a posteriori}.

Once a graph dataset has been obtained, for each adjacency matrix $A^{(k)}$ corresponding to graphs in the dataset, we simulate short-time quantum dynamics and compute the corresponding feature vector $\mathbf{x}^{(k)}$ along with the target observable $\mathcal{O}_k$. The latter are organized in training patterns $e_k=( \mathbf{x}^{(k)},\,\mathcal{O}_k   )$,  collected to form the training set $\{e_k\}_{k=1}^{k_{max}}$. An independent test set is obtained in a similar way.

The readout weights are determined through ridge regression\cite{Rreg}:
\[
\mathbf{w}^* = \arg\min_{\mathbf{w}} \left\{ \sum_k \left( \mathcal{O}_k - \mathbf{w} \cdot \mathbf{x}^{(k)} - b \right)^2 + \lambda \|\mathbf{w}\|^2 \right\},
\]
where $\lambda$ is the regularization parameter. Once the weights are trained, the performance of the model is first evaluated in the training set and subsequently evaluated in a disjoint test set by comparing the predicted values $\hat{\mathcal{O}}_k$ with the true values $\mathcal{O}_k$.

Performance is quantified by the mean absolute percentage error (MAPE),
\[
\text{MAPE} = \frac{100}{N_s} \sum_{k=1}^{N_s} \left| \frac{\hat{\mathcal{O}}_k - \mathcal{O}_k}{\mathcal{O}_k} \right|,
\]
over both training and test sets, where $N_s$ is the size of the dataset. The Pearson correlation coefficient\cite{StatBook} $r=Cov(\mathcal{O},\hat{\mathcal{O}})/(\sigma_{\mathcal{O}} \sigma_{\hat{\mathcal{O}}})$ between $\hat{\mathcal{O}}_k$ and $\mathcal{O}_k$ is also considered to assess the linear agreement between predictions and targets and complement MAPE metrics.

\paragraph*{Performance analysis.---} 
In Table~\ref{tab:results}, we summarize the performance of our protocol over a variety of learning tasks involving global observables of the graph. For each task, the table reports the composition of the training and test sets, the MAPE on both sets and, when necessary, the Pearson correlation coefficient $r$ between the predicted and true values on the training and test sets.

To ensure that all target quantities contribute comparably to the error metrics, each observable has been preprocessed and normalized to a suitable reference value. For instance, spectral moments such as $\mathrm{Tr}[A^k]$ are normalized to the moment of the complete graph\cite{Chung1997} of the same size, i.e. $\mathrm{Tr}[A_c^k]$ with $[A_c]_{ij}=1-\delta_{ij}$. This ensures that all training targets lie within comparable ranges and that regression weights remain numerically balanced.

Several of the quantities of interest have clear topological or spectral significance. In particular, $\mathrm{Tr}[A^2]$ is proportional to the sum of all node degrees and is thus directly related to the average connectivity. $\mathrm{Tr}[A^3]$ counts (six times) the number of triangles in the network, while $\mathrm{Tr}[A^4]$ encodes the total number of closed walks of length four, including quadrilaterals and other small motifs.

The \textit{hub density}, defined as the fraction of nodes whose degree exceeds the average degree by one standard deviation, quantifies the presence of dominant nodes that may act as structural or dynamical bottlenecks. Finally, the spectral ratio $|\lambda_{\min}/\lambda_{\max}|$, constructed from the smallest and largest eigenvalues of the adjacency matrix, provides a measure of the spectral symmetry around zero, a feature that directly affects the coherence and efficiency of excitation transport.

These observables, although nonlocal in nature, are successfully predicted from short-time dynamics restricted to a few monitored sites, emphasizing the information richness of transient quantum evolutions.

Overall, we observe excellent generalization performance across all tasks. Typical test-set MAPE values lie below 10\%, and in many cases below 5\%. When MAPE appears higher (e.g., in learning tasks involving spectral moments such as $\mathrm{Tr}[A^3]$ and $\mathrm{Tr}[A^4]$, which tend to attain small values in Erdős–Rényi graphs with low connection probability $p \lesssim 0.3$), the Pearson correlation coefficient remains close to unity, indicating that the predictive relation is preserved and the error is merely inflated by small target values. This behavior is consistent with the known limitations of relative-error metrics, and underscores the value of using $r$ as a complementary performance indicator.
\\ 
\paragraph*{Intrusion modeling and detection. ---} 
A supervisor monitoring the network and aware of its size $n$ may efficiently estimate it through local probes, as confirmed by the excellent performance obtained in the \textit{network size} task (Tab.~\ref{tab:results}). Deviations from the expected size may then signal the presence of unauthorized connections.
Such a scenario may correspond to a \emph{passive attack}, in which an external system is connected to the monitored network via a single link. A more disruptive situation corresponds to an \emph{active intrusion}, in which the connected device exerts a dynamical action on the network. We adopt a simple model of this condition by introducing a local non-Hermitian term to the Hamiltonian $H = \gamma A$, which becomes
\begin{equation}
    H_\mathrm{eff} = \gamma A - i \Gamma \ket{\alpha}\bra{\alpha} ,
\end{equation}
where $\Gamma > 0$ quantifies the strength of the leakage, and $\ket{\alpha}\bra{\alpha}$ denotes the projector operator on the site $\alpha$ (distinct from the monitored ones) where the parasitic channel is attached. The resulting dynamics is no longer unitary, although the evolution operator can still be written as $U(t)=\exp(-iH_{eff}t/\hbar)$. Thus, the norm of a generic state $\ket{\psi(t)}$ satisfies the equation
\begin{equation}
    \frac{d}{dt} \langle \psi(t)|\psi(t) \rangle = - \frac{2\Gamma}{\hbar} |\langle \alpha|\psi(t) \rangle |^2 ,
\end{equation}
which indicates that loss of normalization occurs only when the system state develops a finite overlap with the intrusion site $\alpha$, implying partial trapping of the excitation by the unmonitored device.

To assess whether the leakage strength $\Gamma$ can be inferred from local measurements, we follow the same formal protocol adopted in the Hermitian case, with the sole modification of accounting for the non-unitary nature of the evolution. 

We construct a dataset denoted by $[G_\Gamma(100, 0.5)]_{360}$ (see Table \ref{tab:results}), consisting of 360 graphs generated by deforming the connectivity of a fixed Erdős–Rényi graph belonging to the ensemble $G(100, 0.5)$ through the addition of an imaginary self-loop of the form $ -i\frac{\Gamma}{\gamma}\,|\alpha\rangle\langle\alpha|$, with $\alpha$ fixed and $\Gamma$ a random variable uniformly distributed in $[0, 2\gamma]$. The resulting graphs in the dataset differ solely in the strength of the imaginary self-loop applied at node $\alpha$. A test set of 40 samples is generated in a similar way.

\begin{figure}
    \centering
    \includegraphics[width=0.98\linewidth]{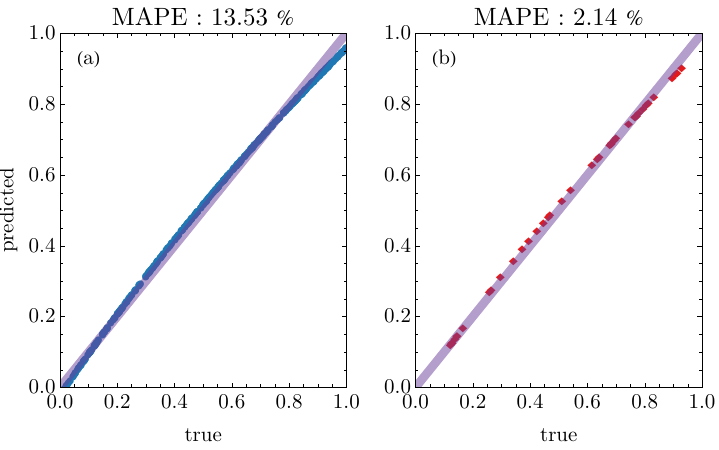}
    \caption{Prediction of the leakage strength parameter $\Gamma$. Each marker denotes one data instance (i.e.\ one sample) for which $\Gamma_{\rm true}$ (horizontal axis) is compared to $\Gamma_{\rm pred}$ (vertical axis). In panel (a) the points refer to the training set, while in panel (b) to the test set. The solid diagonal line indicates the ideal relation $\Gamma_{\rm pred} = \Gamma_{\rm true}$. The mean absolute percentage error (MAPE) is $13.53\%$ on the training set and $2.14\%$ on the test set. The relative high MAPE in training is mainly due to larger relative errors for very small and very large values of $\Gamma$, which are absent in the test set, thus resulting in a more accurate performance.}
    \label{fig:2}
\end{figure}

Despite the non-conservative nature of the dynamics, the inference method remains accurate. In particular, the Pearson correlation on the test set exceeds $r = 0.999$, demonstrating that the strength of the non-Hermitian coupling, here used as a marker of information leakage, can be effectively deduced from local, short-time measurements. This conclusion is further confirmed by the true-predicted plots in Fig.~\ref{fig:2}, showing consistent performance on both training and test sets, with larger but still limited deviations in the training set at very small and very large values of $\Gamma$. Remarkably, the impact of the non-Hermitian deformation is detectable well before any significant loss of state normalization occurs, as the spectral properties of the system are profoundly modified by the local imaginary potential\cite{NHermit}. These findings support the use of deviations from unitarity as early, local diagnostic signals for intrusion detection, with potential implications in quantum cybersecurity frameworks.

\paragraph*{Scalability and Noise Resilience of the Protocol.---}
Up to this point, we have discussed the protocol under ideal conditions, implicitly assuming that local occupation probabilities can be determined with arbitrary precision. However, in realistic quantum systems, probabilities are obtained through repeated measurements, so that statistical fluctuations, imperfect detection, and signal losses inevitably affect the estimates. It is therefore essential to assess the resilience of the protocol to such fluctuations, as well as its scalability with system size.

\begin{figure}
    \centering
    \includegraphics[width=0.98\linewidth]{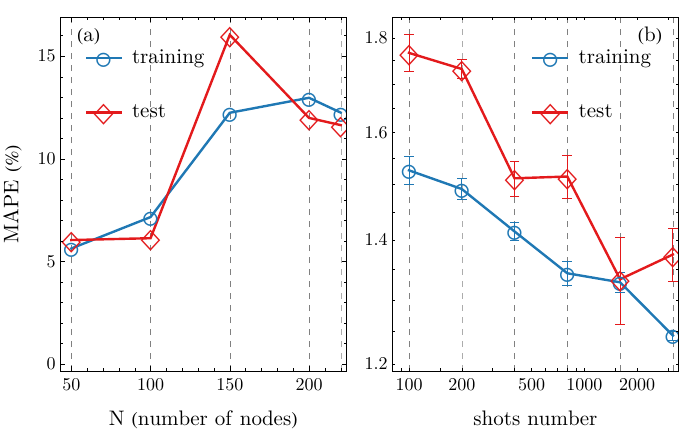}
    \caption{Robustness analysis of the protocol. (a) Mean absolute percentage error (MAPE) for the prediction of $\mathrm{Tr}[A^2]$ (third row of Table~\ref{tab:results}) as a function of the number of graph nodes $N$, while keeping fixed the number of monitored sites ($5$). As expected, the performance slightly deteriorates with increasing $N$, yet remains accurate up to $N \approx 230$. (b) MAPE for the prediction of $\mathrm{Tr}[A^2]$ (second row of Table~\ref{tab:results}) when the theoretical occupation probabilities are perturbed with Gaussian fluctuations of zero mean and the standard deviation $\sqrt{p(1-p)/N_{\text{shot}}}$, with $p$ the theoretical probability and $N_{\text{shot}}$ the number of repeated measurements. Data points and error bars show the mean and standard deviation of the mean over six independent realizations. The performance is already close to the ideal limit for relatively small $N_{\text{shot}}$, and converges to it for $N_{\text{shot}} \sim 3000$.}
    \label{fig:3}
\end{figure}

Panel (a) of Fig.~\ref{fig:3} addresses the latter point by focusing on the prediction of $\mathrm{Tr}[A^2]$ (third row of Table~\ref{tab:results}) while systematically increasing the number of graph nodes $N$, without altering the number of monitored sites (fixed to five). Since larger networks would in principle require access to a larger subset of nodes to retain predictive power, a degradation of performance at fixed monitoring size is expected. This effect is indeed observed: the mean absolute percentage error (MAPE) increases moderately with $N$, yet the overall accuracy remains high up to the largest tested networks ($N= 220$). These results show that the protocol maintains good predictive capability even when the monitored fraction of the system becomes vanishingly small.
\\ 
Panel (b) of Fig.~\ref{fig:3} examines the robustness against statistical noise in the estimation of local occupation probabilities. Starting from the task of predicting $\mathrm{Tr}[A^2]$ (second row of Table~\ref{tab:results}), the theoretical probabilities were perturbed by Gaussian fluctuations of zero mean and standard deviation $\sqrt{p(1-p)/N_{\text{shot}}}$, with $p$ the theoretical probability and $N_{\text{shot}}$ the number of repeated measurements. For each $N_{\text{shot}}$, the mean and the standard deviation of the mean of performance metrics in six independent realizations are shown as data points with error bars. The performance is already close to the ideal case with a relatively small number of measurements and approaches the ideal limit for $N_{\text{shot}}\sim 3000$. Additional sources of uncertainty, such as signal losses or imperfect detection, effectively induce similar stochastic fluctuations in the estimated probabilities and therefore impact the performance in a qualitatively analogous way.\\
These results demonstrate that the protocol maintains high accuracy at realistic noise levels and moderate network scaling, supporting its applicability for experimental implementation.

\paragraph*{Conclusion.---} 
We have shown that key global properties of a quantum network, such as degree statistics, cycle structure, and signatures of non-Hermiticity, can be accurately inferred from strictly local quantum measurements. Our protocol combines the expressivity of short-time quantum dynamics with the simplicity of a non-iterative linear readout, in a scheme reminiscent of quantum reservoir computing and extreme learning. These results reveal the intrinsic fragility of networked quantum systems, in which sensitive structural information can be deduced by unauthorized local measurements. At the same time, they point to novel strategies for network monitoring and security diagnostics in quantum infrastructures.\\
The extension of the present framework to initial states involving two or more quantum excitations\cite{QWPhotons1} represents an important avenue for future research. Such generalizations may unlock new functionalities or enhance the performance of existing tasks, by leveraging multi-particle interference and entanglement in the inference process.\\
Beyond the scope of quantum network diagnostics, our approach may also inspire novel quantum algorithms for image recognition, pattern analysis, and non-gate-based quantum computation.
\paragraph*{Acknowledgments.---} 
\begin{acknowledgments}
The research activity of FR received support from the PNRR MUR project PE0000023 – NQSTI, through the cascade-funded projects SPUNTO (CUP E63C22002180006, D.D. MUR No. 1564/2022) and TOPQIN (CUP J13C22000680006, D.D. MUR No. 1243/2022).\\
JS acknowledges the contribution from PRIN (Progetti di Rilevante Interesse Nazionale) TURBIMECS, grant n. 2022S3RSCT CUP H53D23001630006, and PNRR MUR project PE0000023 - NQSTI through the secondary projects “ThAnQ” J13C22000680006.\\ 
\end{acknowledgments}

\bibliography{refs}

\end{document}